# Surface-bulk coupling in a Bi$_2$Te$_3$ nanoplate grown by van der Waals epitaxy

Xiaobo Li[1,2], Mengmeng Meng[1], Shaoyun Huang[1], Congwei Tan[3], Congcong Zhang[3], Hailin Peng[3], and H. Q. Xu[1,4,*]

[1]*Beijing Key Laboratory of Quantum Devices, Key Laboratory for the Physics and Chemistry of Nanodevices, and School of Electronics, Peking University, Beijing 100871, China*

[2] *Academy for Advanced Interdisciplinary Studies, Peking University, Beijing 100871, China*

[3] *Center for Nanochemistry, Beijing National Laboratory for Molecular Sciences (BNLMS), College of Chemistry and Molecular Engineering, Peking University, Beijing 100871, China*

[4] *Beijing Academy of Quantum Information Sciences, Beijing 100193, China*

[*]Corresponding authors: H. Q. Xu (hqxu@pku.edu.cn)

(Date: January 19, 2022)

## ABSTRACT

We report on an experimental study of the effect of coherent surface-bulk electron scattering on quantum transport in a three-dimensional topological insulator Bi$_2$Te$_3$ nanoplate. The nanoplate is grown via van der Waals epitaxy on a mica substrate and a top-gated Hall-bar device is fabricated from the nanoplate directly on the growth substrate. Top-gate voltage dependent measurements of the sheet resistance of the device reveal that the transport carriers in the nanoplate are of n-type and that, with decreasing top gate voltage, the carrier density in the nanoplate is decreased. However, the mobility is increased with decreasing top-gate voltage. This mobility increase with decreasing carrier density in the nanoplate is demonstrated to arise from a decrease in bulk-to-surface electron scattering rate. Low-field magnetotransport measurements are performed at low temperatures. The measured magnetoconductivity of the nanoplate shows typical weak anti-localization (WAL) characteristics. We analyze the measurements by taking surface-bulk inter-channel electron scattering into account and extract dephasing times $\tau_\varphi$, diffusion coefficients $D$ of electrons at the top surface and in the bulk, and the surface-bulk scattering times $\tau_{SB}$ as a function of top-gate voltage and temperature. It is found that the dephasing in the nanoplate arises dominantly from electron-electron scattering with small energy transfers. It is also found that the ratio of $\tau_\varphi/\tau_{SB}$ (a measure of the surface-bulk electron coherent coupling) is decreased with decreasing gate voltage or increasing temperature. We demonstrate that taking the surface-bulk coherent electron scattering in our $Bi_2Te_3$ nanoplate into account is essential to understand quantum transport measurements at low temperatures.

Three-dimensional (3D) topological insulators (TIs) are a new class of quantum materials which possess gapped bulk and gapless surface states.[1, 2] Due to time-reversal symmetry and spin-momentum interlock, these surface, topological, helical Dirac fermion states are attractive for applications in low energy-dissipative spintronics[3,4] and for construction of Majorana bound states.[5] In physics, the helical Dirac fermion quasiparticles at the surface traveling through the coherent time-reversed closed paths can interfere, leading to a positive quantum correction to the conductivity and thus the weak anti-localization (WAL) effect.[6] In experiments, the detection of topological surface states in 3D TIs is usually carried out through analysis of the WAL effect observed in transport measurements at low magnetic fields.[7-9] Unfortunately, due to strong spin-orbit coupling, the conducting bulk state of a 3D TI can also contribute to the WAL effect, which masks the surface signals and thus makes the surface state analysis difficult.[10,11] To extract the surface state contribution from transport measurements in a 3D TI, one usually carries out the transport measurements of the material at different carrier densities achieved via electrostatic tuning and analyzes the quantum correction to the magnetoconductivity based on the Hikami-Larkin-Nagaoka (HLN) formula.[12-14] Through fits of the magnetoconductivity to the HLN formula, it has been demonstrated that the extracted absolute value of dimensionless prefactor $|\alpha|$ in the formula (a measure of the number of 2D conduction channels in the material) usually varies from ~1/2 (surface-bulk state coupled transport in the device) to ~1 (transport through separate surface and bulk channels),[15-17] reflecting the existence of the surface states. However, since the effect of the surface-bulk electron scattering is not included in the HLN formula, analysis of the magnetoconductivity based on this formula could make deviations, especially for a case with a strong surface-bulk state coupling. Garate and Glazman have evaluated the quantum correction to magnetoconductivity in a 3D TI film with surface-bulk scattering being taken into consideration.[18] It has been shown that, once states at a surface of a 3D TI film are coupled to the bulk states, the quantum correction to the magnetoconductivity should in general be expressed as a function of dephasing times $\tau_\varphi$ and diffusion coefficients $D$ of carriers at the surface and in the bulk, and surface-bulk inter-channel scattering time $\tau_{SB}$. Using this model to analyze the measured magnetoconductivity, one can extract the transport parameters $\tau_\varphi$, $\tau_{SB}$ and $D$ in a 3D TI film, as well as the ratio of $\tau_\varphi/\tau_{SB}$, i.e., a measure of the strength of the surface-bulk state coupling (with $\tau_\varphi/\tau_{SB} > 1$ referring to the surface-bulk-coherently-coupled case and $\tau_\varphi/\tau_{SB} < 1$ referring to the case where the surface and bulk states are no longer coherently coupled).[15,19]

Here, we report on a low-temperature transport study of a $Bi_2Te_3$ nanoplate. The $Bi_2Te_3$ nanoplate is grown on a mica substrate via van der Waals epitaxy and the transport measurements are made in a top-gated Hall-bar device setup. Top-gate voltage dependent measurements of the sheet resistance of the nanoplate reveal top surface-bulk state coupled, n-type carrier transport characteristics. Low-field magnetotransport measurements are performed for the $Bi_2Te_3$ nanoplate Hall-bar device at different gate voltages and temperatures. The dephasing times $\tau_\varphi$ and the diffusion coefficients $D$ of electrons at the surface and in the bulk, and the surface-bulk electron scattering times $\tau_{SB}$ are extracted by fits of the measured magnetoconductivity to the Garate-Glazman theory. It is found that the ratio of $\tau_\varphi/\tau_{SB}$ is gate-voltage and temperature dependent. As the gate voltage decreases or the temperature increases, $\tau_\varphi/\tau_{SB}$ is decreased and thus the strength of the surface-bulk state coupling is weakened.

$Bi_2Te_3$ nanoplates are grown on a mica substrate with a fresh, cleavage plane via van der Waals epitaxy.[20] The inset of Fig. 1(a) shows an optical image of as-grown $Bi_2Te_3$ nanoplates. These nanoplates have a lateral size of 10 ~ 30 μm and a thickness of 4 ~ 20 nm. In fabrication of a Hall-bar device, metal makers are first defined on the growth substrate to locate nanoplates. Then, six electrodes (source, drain and four probe contacts) consisting of 5-nm-thick Ti and 90-nm-thick Au are made on a selected nanoplate by electron-beam lithography for pattern definition, electron-beam evaporation for metal deposition, and lift-off process. After that, a dielectric layer of 20-nm-thick $HfO_2$, covering the entire Hall bar and sufficiently large portions of contact electrode, is made by electron-beam lithography, atomic layer deposition and lift-off. Finally, a top gate made of 5-nm-thick Ti and 90-nm-thick Au is fabricated on the $HfO_2$ dielectric layer again by electron-beam lithography, electron-beam evaporation and lift-off process. Figure 1(a) shows an optical image of a fabricated top-gated Hall-bar device. The device is made from a hexagonal $Bi_2Te_3$ nanoplate [as marked by the white arrow in the inset of Fig. 1(a)] with a thickness of ~12 nm.

Figure 1(b) shows a schematic diagram of the device structure and the measurement circuit setup. The distance $L$ between probes 1 and 2 or between probes 3 and 4 is designed as ~6 μm, and the width $W$ of the Hall bar (i.e., the measured distance between probes 1 and 3 or between probes 2 and 4) is designed as ~20 μm. Cryogenic transport measurements are performed in a physical property measurement system (PPMS) equipped with a superconducting magnet which can provide a magnetic field up to 9 T. In the measurements, both longitudinal voltage $V_{xx}$ and transverse voltage $V_{yx}$ are detected using a standard lock-in technique, in which an ac current $I$ of 300 nA at a frequency of 17 Hz is supplied between the source and drain contacts. The longitudinal resistance $R_{xx}$ and the Hall resistance $R_{yx}$ are obtained numerically from the measurements as $R_{xx} = \frac{V_{xx}}{I}$ and $R_{yx} = \frac{V_{yx}}{I}$.

Figure 1(c) shows the top-gate voltage dependent measurements of the sheet resistance $R$ (defined as $R = R_{xx} \times \frac{W}{L}$) of the device shown in Fig. 1(a) at $T = 2$ K. It is seen that $R$ increases as the top gate voltage decreases, indicating that the transport carriers in the nanoplate are of electrons. Previously, n-type $Bi_2Te_3$ layers synthesized by means of van der Waals epitaxy[21-23] have been reported and the n-type carriers in the materials are thought to be originated from Bi vacancies as well as $Te_{Bi}$ antisites.[24,25] The blue vertical dashed line in Fig. 1(c) marks the top-gate voltage value of $V_g = 5$ V. As $V_g$ decreases from $10$ V to $5$ V, the depletion of electrons in the nanoplate causes an increase of $R$. The sheet resistance $R$ becomes to increase more slowly as the top-gate voltage continues to decrease from $V_g = 5$ V. This could presumably be considered as a result of enhanced screening against the top gate by gradually separated top-surface states. To study the coupling effect between the top surface and bulk states at different carrier densities, we carry out the magnetotransport measurements at $V_g = -10$ V, $-3$ V, 0 V, 5 V and 11 V. The inset of Fig. 1(c) shows the sheet electron density $n$, together with the Hall mobility $\mu$, extracted from the measurements. It is found that $n$ decreases as $V_g$ decreases and reaches to a value of $\sim 2.5 \times 10^{14}$ cm$^{-2}$ at $V_g = -10$ V. The mobility $\mu$ is found to increase as $V_g$ decreases, which implies that transport scattering becomes weakened with decreasing $V_g$ (we will further discuss this important point later). Figure 1(d) shows detailed measurements of the longitudinal resistance $R_{xx}$ and the Hall resistance $R_{yx}$ of the device as a function of the magnetic field $B$ at $T = 2$ K and $V_g = 0$ V. The magnetic field is applied perpendicular to the substrate. The sharp dip seen in the longitudinal resistance $R_{xx}$ in the vicinity of $B = 0$ T reveals a typical WAL effect. The negative slope of the Hall resistance $R_{yx}$ seen in Fig. 1(d) again indicates that the transport carriers in the nanoplate are of electrons.

Figure 2(a) shows the measured magnetoconductivity, $\Delta\sigma_{xx}(B) = \sigma_{xx}(B) - \sigma_{xx}(B = 0)$, at the top-gate voltages selected above and $T = 2$ K (opened circles). Here, $\sigma_{xx}$ is obtained from $\sigma_{xx} = \frac{R_{xx}}{R_{xx}^2 + R_{yx}^2} \times \frac{L}{W}$ and the measured curves are successively vertically offset for clarify. These magnetoconductivity curves all show typical WAL characteristics. We could analyze these observed WAL characteristics using the Hikami-Larkin-Nagaoka (HLN) formula[26] as it is commonly done in the literature. However, the extracted temperature dependence of the dephasing length from such analyses has been found to deviate strongly from one that would be expected for a 2D system (see Supplementary Materials). Here we are going to analyze our measured WAL characteristics based on a formula derived previously by Garate and Glazman[18], in which the electron transport at the surface and in the bulk, as well as electron scattering between the surface and the bulk states, have been treated explicitly. In the case for our top-gated $Bi_2Te_3$ nanoplate device, we consider that the top surface and the bulk are the two 2D conduction channels and assume that the bottom surface has a negligible contribution to the 2D WAL effect. This assumption is justified by the fact that it has been hard to observe a contribution to the quantum correction to the conductivity from the bottom surface in a 3D topological insulator grown on a mica substrate[27,28], due to the presence of a substantial concentration of defects or stacking faults in the first few layers grown on the mica substrate[29]. In the view of that the top surface and the bulk of the $Bi_2Te_3$ nanoplate are the two 2D conduction channels and by considering the electron scattering between the top surface and bulk states, the Garate-Glazman formula for the quantum correction to the magnetoconductivity is given by[18]

$$\Delta\sigma_{xx}(B) = \frac{e^2}{2\pi h}\left[f\left(\frac{\hbar q_+^2}{4eB}\right) + f\left(\frac{\hbar q_-^2}{4eB}\right)\right], \tag{1}$$

where $f(x) = ln(x) - \psi(1/2 + x)$, with $\psi$ being the digamma function, and $q_\pm^2 = \frac{1}{2}(\frac{1}{\tilde{l}_{\varphi_1}^2} + \frac{1}{\tilde{l}_{\varphi_2}^2} \pm \sqrt{(\frac{1}{\tilde{l}_{\varphi_1}^2} - \frac{1}{\tilde{l}_{\varphi_2}^2})^2 + \frac{4}{l_{SB_1}^2 l_{SB_2}^2}})$, with $\tilde{l}_{\varphi_i} = \sqrt{D_i \tilde{\tau}_{\varphi_i}}$, $\tilde{\tau}_{\varphi_i}^{-1} = \tau_{\varphi_i}^{-1} + \tau_{SB_i}^{-1}$, $l_{SB_i} = \sqrt{D_i \tau_{SB_i}}$ and i = 1 or 2. Here, $\tau_{\varphi_i}$ and $D_i$ are the dephasing time and diffusion coefficient of electrons in the bulk (i = 1) and at the surface (i = 2), $\tau_{SB_1}$ is the electron scattering time from the bulk to the surface, while $\tau_{SB_2}$ is the electron scattering time from the surface to the bulk.

We fit the measured magnetoconductivity data at the different gate voltages to Eq. (1). The dashed lines in Fig. 2(a) show the results of the fits. It is seen that all the measured magnetoconductivity data can be fitted excellently by Eq. (1). Figures 2(b) and 2(c) show the extracted $\tau_{\varphi_1}$ and $\tau_{\varphi_2}$, $\tau_{SB_1}$ and $\tau_{SB_2}$, and $D_1$ and $D_2$ from the fits. Comparing dephasing times $\tau_{\varphi_1}$ and $\tau_{\varphi_2}$, we find that $\tau_{\varphi_2}$ is much longer than $\tau_{\varphi_1}$ at all the selected top-gate voltages. Here, we note that since the top-surface state electrons are topologically protected, it is reasonable to assign $\tau_{\varphi_2}$ to the dephasing time of electrons at the surface. In addition, we find that both $\tau_{\varphi_1}$ and $\tau_{\varphi_2}$ decrease as the top-gate voltage decreases. These gate voltage dependences of $\tau_{\varphi_1}$ and $\tau_{\varphi_2}$ indicate a stronger dephasing process at a lower top-gate voltage, due to a reduction in Coulomb screening and an enhancement in electron-electron interaction[30] at a lower electron density. The bulk-to-surface and surface-to-bulk electron scattering times show different top-gate voltage dependences: $\tau_{SB_1}$ increases as the top-gate voltage decreases, while $\tau_{SB_2}$ remains nearly unchanged. These results are excellently in agreement with the fact that the surface-to-bulk scattering rate is proportional to the density of states (DOS) in the bulk, i.e., $\tau_{SB_2}^{-1} \propto$ DOS in the 2D bulk, which is a constant, while the bulk-to-surface scattering rate is proportional to the DOS at the surface, i.e., $\tau_{SB_1}^{-1} \propto$ DOS at the top surface, which increases with increasing Fermi energy (see Supplementary Materials). It is now important to note that the observed increase in the Hall mobility with decreasing gate voltage, as we showed and discussed above, could be attributed to the fact that the inter-2D-channel scattering time $\tau_{SB_1}$ of electrons in the bulk is significantly increased as $V_g$ decreases. In addition, $\tau_{SB_1}$ is found to be 2.2~3.9 times larger than $\tau_{SB_2}$, which indicates that the bulk possesses a higher DOS than the top surface at the considered values of $V_g$. The diffusion coefficients extracted for the bulk $D_1$ and for the top surface $D_2$ are found to be $D_1 < D_2$ at all these considered top-gate voltages [see the insets in Figs. 2(b) and 2(c)]. It is also seen that both $D_1$ and $D_2$ are weakly dependent on $V_g$ [see again the insets in Figs. 2(b) and 2(c)], indicating that the electron diffusion is hardly affected by the electron density at these top-gate voltages. The weak $V_g$ dependences of the diffusion coefficients are consistent with the expectation that $D_1, D_2 \propto E_F \mu$ (see Supplementary Materials) and the observation shown in the inset of Fig. 1(c) that $\mu$ increases almost linearly with decreasing $V_g$ (note that, to a good approximation, the Fermi energy $E_F$ can be assumed to decrease linearly with $V_g$). This result shows again the importance of considering surface-bulk scattering in analyzing the transport properties of the $Bi_2Te_3$ nanoplate. Figure 2(d) shows the dephasing lengths in the bulk $L_{\varphi_1}$ and at the top surface $L_{\varphi_2}$ extracted via $L_{\varphi_1} = \sqrt{D_1 \tau_{\varphi_1}}$ and $L_{\varphi_2} = \sqrt{D_2 \tau_{\varphi_2}}$. As expected, $L_{\varphi_2}$ exhibits a longer dephasing length than $L_{\varphi_1}$ and both become smaller at a lower electron density.

To quantify the influence of the top-gate voltage on the strength of the surface-bulk state coupling, we show in Figs. 2(e) and 2(f) the top-gate voltage dependences of $\tau_{\varphi_1}/\tau_{SB_1}$ and $\tau_{\varphi_2}/\tau_{SB_2}$. Previously, it was shown that the ratio of $\tau_\varphi/\tau_{SB}$ can be used to characterize the effect of the surface-bulk state coupling.[15,19] When $\tau_\varphi/\tau_{SB} > 1$, electrons maintain phase coherence during transferring from one 2D conduction channel to the other 2D channel, leading to the formation of a two-channel-coherently-coupled conduction channel, and the system is in a strong inter-channel coupling case. However, when $\tau_\varphi/\tau_{SB} < 1$, electrons will lose phase coherence before being scattered into the other channel and electrons in the two conduction channels are no longer coherently coupled. Figures 2(e) and 2(f) show that both $\tau_{\varphi_1}/\tau_{SB_1} > 1$ and $\tau_{\varphi_2}/\tau_{SB_2} > 1$ at the considered top-gate voltages. Thus, the electron transport in the $Bi_2Te_3$ nanoplate is in a surface-bulk coherently coupled case. It is also seen in Figs. 2(e) and 2(f) that $\tau_{\varphi_1}/\tau_{SB_1}$ and $\tau_{\varphi_2}/\tau_{SB_2}$ decrease with decreasing $V_g$, reflecting that the strength of the surface-bulk coherent coupling is gradually weakened.

To examine how the transport characteristic parameters in the $Bi_2Te_3$ nanoplate depend on temperature, we show in Fig. 3(a) the measured magnetoconductivity $\Delta\sigma_{xx}$ (opened circles) at a fixed top-gate voltage of $V_g = -10$ V at different temperatures. Here, again, the measured magnetoconductivity data at different temperatures are successively vertically offset for clarify. It is seen that a WAL magnetoconductivity peak appears in all the considered temperatures but is gradually suppressed with increasing temperature. The measured magnetoconductivity data at different temperatures are fitted to Eq. (1) and the results of the fits are presented by the dashed lines in Fig. 3(a). Figures 3(b) and 3(c) show the characteristic transport parameters $\tau_{\varphi_1}$ and $\tau_{\varphi_2}$, $\tau_{SB_1}$ and $\tau_{SB_2}$, and $D_1$ and $D_2$ extracted from the fits. It is found that $\tau_{SB_1}$, $\tau_{SB_2}$, $D_1$ and $D_2$ all are weakly temperature dependent. However, $\tau_{\varphi_1}$ and $\tau_{\varphi_2}$ are found to be strongly temperature dependent and, as expected, they both decrease with increasing temperature. Figure 3(d) shows dephasing lengths $L_{\varphi_1}$ and $L_{\varphi_2}$ obtained from $L_{\varphi_i} = \sqrt{D_i \tau_{\varphi_i}}$ with $i = 1$ and 2. Both $L_{\varphi_1}$ and $L_{\varphi_2}$ are found to exhibit a power-law temperature dependence, $L_{\varphi_1} \sim T^{-0.55}$ and $L_{\varphi_2} \sim T^{-0.49}$, showing that the electron dephasing both at the top surface and in the bulk arises dominantly from electron-electron scattering with small-energy transfers.[30] Figures 3(e) and 3(f) show the temperature dependences of $\tau_{\varphi_1}/\tau_{SB_1}$ and $\tau_{\varphi_2}/\tau_{SB_2}$. It is seen that both $\tau_{\varphi_1}/\tau_{SB_1}$ and $\tau_{\varphi_2}/\tau_{SB_2}$ decrease with increasing temperature (i.e., an increase in temperature will weaken the strength of the surface-bulk state coherent coupling due to a decrease in dephasing time). It is important to note that $\tau_{\varphi_1}/\tau_{SB_1} < 1$ when $T \geq 6$ K. This indicates that electrons in the bulk would lose phase coherence before being scattered into the top surface at $T \geq 6$ K and thus the coherent electron transport in the bulk could well be treated within the single 2D channel framework. However, $\tau_{\varphi_2}/\tau_{SB_2} > 1$ at all considered temperatures, which indicates that the coherent electron transport at the top surface could not be properly treated without considering electron scattering into the bulk at these temperatures.

In conclusion, a top-gated Hall-bar device has been fabricated from a van der Waals epitaxially grown $Bi_2Te_3$ nanoplate directly on the growth mica substrate. The top-gate voltage dependent measurements of the sheet resistance reveal that the transport carriers in the nanoplate are of n-type. The measurements also show that with decreasing top-gate voltage $V_g$, the sheet resistance initially increases quickly at large positive $V_g$ but then turns to exhibit a slower increase with decreasing $V_g$, indicating an enhanced Coulomb screening against the top gate by the gradually separated top surface-state electrons. The carrier density is found to decrease with decreasing $V_g$, but the mobility shows a continuous increase with decreasing $V_g$, implying a reduction in the surface-bulk inter-channel scattering at a lower $V_g$. The measured magnetoconductivity shows the WAL characteristics and is analyzed by taking electron scattering between the top surface and bulk states into account. The dephasing times, $\tau_{\varphi_1}$ and $\tau_{\varphi_2}$, and diffusion coefficients, $D_1$ and $D_2$, of electrons in the bulk and at the top surface, as well as the scattering time $\tau_{SB_1}$ for electron from the bulk to the top surface and the scattering time $\tau_{SB_2}$ for electron from the top surface to the bulk are extracted. It is shown that both $\tau_{\varphi_1}$ and $\tau_{\varphi_2}$ decrease with decreasing $V_g$. However, $\tau_{SB_1}$ is seen to increase with decreasing $V_g$, while $\tau_{SB_2}$ is found to be nearly independent of $V_g$. This result is explained by considering the characteristic difference in the density of states in the bulk and at the top surface. The diffusion coefficients $D_1$ and $D_2$ in the bulk and at the top surface are found to be top-gate voltage $V_g$-independent. This result, together with the observed decrease in $\tau_{\varphi_1}$ and $\tau_{\varphi_2}$ with decreasing $V_g$, leads to the observation that the dephasing lengths in the bulk $L_{\varphi_1}$ and at the top surface $L_{\varphi_2}$ are decreased with decreasing $V_g$, in agreement with the results reported in the literature. The measures of coherent coupling of the bulk and top-surface 2D channels, $\tau_{\varphi_1}/\tau_{SB_1}$ and $\tau_{\varphi_2}/\tau_{SB_2}$ are also extracted. It is found that both increase with increasing $V_g$ and are $>1$ at the considered values of $V_g$ at $T = 2$ K, providing an evidence that the transport is in the top surface-bulk coherently coupled regime.

Temperature-dependent measurements of the magnetoconductivity in the top-gated $Bi_2Te_3$ nanoplate Hall-bar device are also performed at $V_g = -10$ V (a weak top surface-bulk coherently coupled case). It is found that for temperatures $T = 2$ K to 14 K, the diffusion coefficients, $D_1$ and $D_2$, and the inter-channel scattering times, $\tau_{SB_1}$ and $\tau_{SB_2}$, all show to be temperature independent. However, the dephasing times $\tau_{\varphi_1}$ and $\tau_{\varphi_2}$, the dephasing lengths $L_{\varphi_1}$ and $L_{\varphi_2}$, and the ratios $\tau_{\varphi_1}/\tau_{SB_1}$ and $\tau_{\varphi_2}/\tau_{SB_2}$ are all found to be decreased with increasing temperature. Importantly, it is found that $L_{\varphi_1} \sim T^{-0.55}$ and $L_{\varphi_2} \sim T^{-0.49}$, showing that the electron dephasing is dominated by the electron-electron scattering with small energy transfers. In addition, $\tau_{\varphi_1}/\tau_{SB_1}$ is found to become $< 1$ at temperatures $T = 6$ K to 14 K. This result shows that the transport in the bulk can be treated as coherently decoupled from the top surface. However, $\tau_{\varphi_2}/\tau_{SB_2}$ is found to be $> 1$ at $T = 2$ K to 14 K, showing that the transport in our $Bi_2Te_3$ nanoplate cannot be strictly treated as coherent transport through fully decoupled top-surface and bulk 2D channels at these low temperatures. Our study presented in this work shows that it is important to consider surface-bulk state coupling explicitly when analyzing the magnetotransport in a 3D topological insulator nanoplate.

## AUTHOR CONTRIBUTIONS

H. Q. Xu conceived and supervised the project. X. Li grew the material, fabricated the device and performed the transport measurements. X. Li and H. Q. Xu analyzed the measurement data. M. Meng and S. Huang participated in the device fabrication, the transport measurements, and the data analysis. C. Tan, C. Zhang and H. Peng participated in the material growth and performed the atomic force microscopy (AFM) measurements of the material. X. Li and H. Q. Xu wrote the manuscript with contributions from all the authors.

## CONFLICTS OF INTEREST

There are no conflicts of interests to declare.

## ACKNOWLEDGEMENTS

This work is supported by the Ministry of Science and Technology of China through the National Key Research and Development Program of China (Grant Nos. 2017YFA0303304, 2017YFA0204901, 2016YFA0300601, and 2016YFA0300802), the National Natural Science Foundation of China (Grant Nos. 92165208, 11874071, 91221202, 91421303, and 11974030), the Beijing Natural Science Foundation (Grant No. 1202010), and the Beijing Academy of Quantum Information Sciences (No. Y18G22).

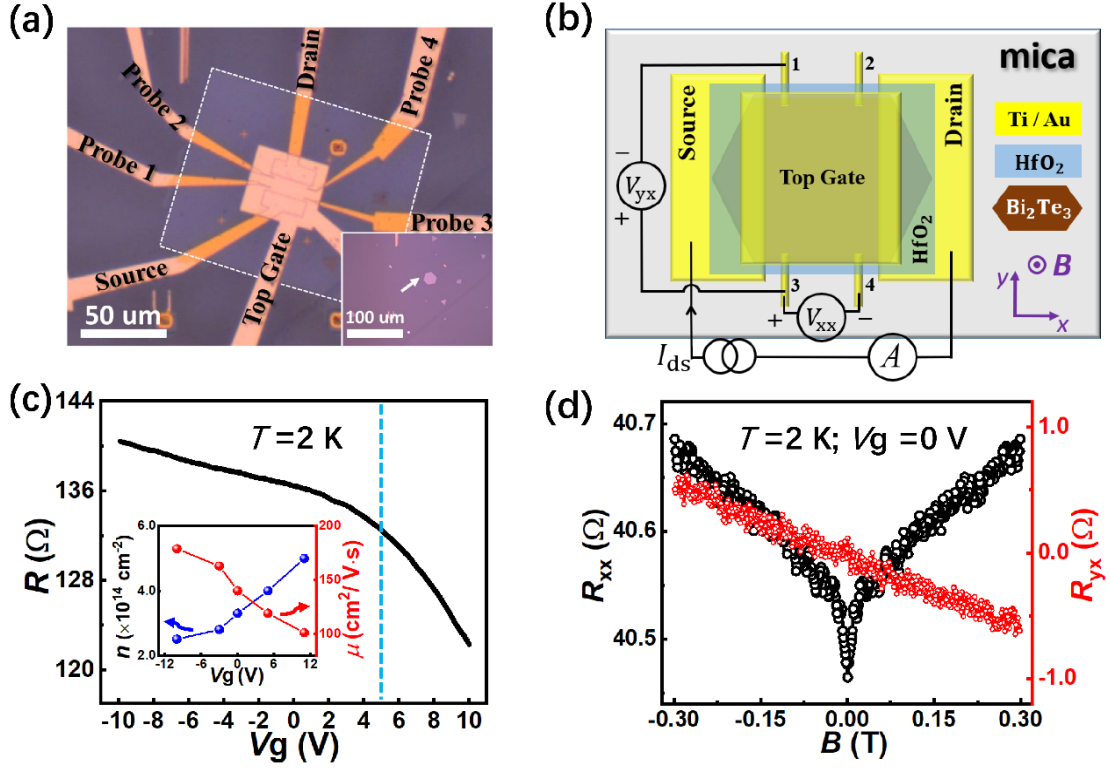

FIG. 1. (a) Optical image of the top-gated Hall-bar $Bi_2Te_3$ nanoplate device studied in this work. The inset shows an optical image of a few $Bi_2Te_3$ nanoplates grown on mica. The device is made from a ~12-nm-thick, hexagonal nanoplate marked by the white arrow in the inset. (b) Schematic diagram of the device structure and the measurement circuit setup. (c) Measured sheet resistance $R$ of the device vs. top-gate voltage $V_g$ at $T = 2$ K. The blue vertical dashed line marks a top-gate voltage value of $V_g = 5$ V. Inset: measured sheet electron density $n$ and Hall mobility $\mu$ vs. top-gate voltage $V_g$ at $T = 2$ K. (d) Longitudinal resistance $R_{xx}$ and Hall resistance $R_{yx}$ of the device as a function of the magnetic field $B$ at $T = 2$ K and $V_g = 0$ V. The magnetic field $B$ is applied perpendicular to the substrate, as indicated in Fig 1(b).

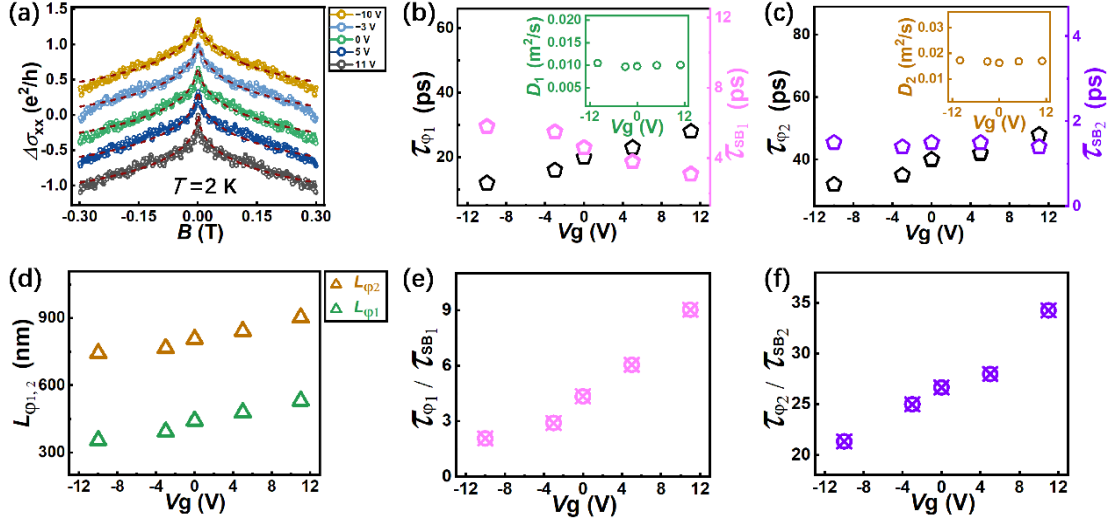

FIG. 2. (a) Magnetoconductivity $\Delta\sigma_{xx}$ of the device at different top gate voltages at $T = 2$ K. The opened circles show the measured data. The black ones are obtained at $V_g = 11$ V and the ones at other values of $V_g$ are successively vertically offset for clarify. The dashed lines show the fits of the measured data using Eq (1). (b) Dephasing time $\tau_{\varphi_1}$ and bulk-to-surface scattering time $\tau_{SB_1}$ of electrons in the bulk vs. top-gate voltage $V_g$. Inset: diffusion coefficient $D_1$ of electrons in the bulk vs. top-gate voltage $V_g$. (c) Dephasing time $\tau_{\varphi_2}$ and surface-to-bulk scattering time $\tau_{SB_2}$ of electrons at the surface vs. top-gate voltage $V_g$. Inset: diffusion coefficient $D_2$ of electrons at the surface vs. top-gate voltage $V_g$. (d) Dephasing lengths $L_{\varphi_1}$ and $L_{\varphi_2}$ vs. top-gate voltage $V_g$. (e) $\tau_{\varphi_1}/\tau_{SB_1}$ vs. top-gate voltage $V_g$. (f) $\tau_{\varphi_2}/\tau_{SB_2}$ vs. top-gate voltage $V_g$.

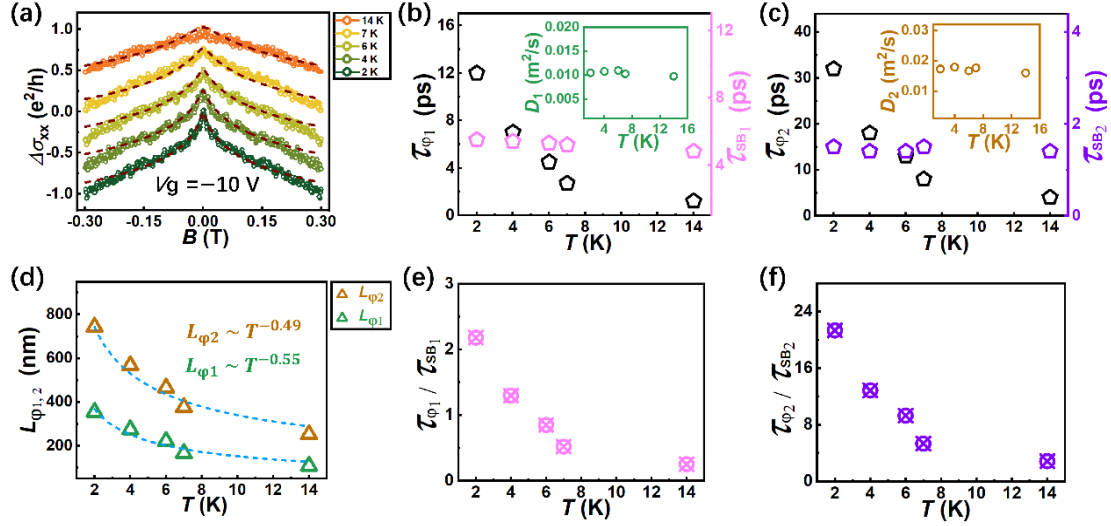

FIG. 3. (a) Magnetoconductivity $\Delta\sigma_{xx}$ of the device at different temperatures at $V_g = -10$ V. Again, the opened circles show the measured data. The dark green ones are obtained at temperature $T = 2$ K and the ones at other temperatures are successively vertically offset for clarify. The dashed lines show the fits of the measured data using Eq (1). (b) Dephasing time $\tau_{\varphi_1}$ and bulk-to-surface scattering time $\tau_{SB_1}$ of electrons in the bulk vs. temperature $T$. Inset: diffusion coefficient $D_1$ of electrons in the bulk vs. temperature $T$. (c) Dephasing time $\tau_{\varphi_2}$ and surface-to-bulk scattering time $\tau_{SB_2}$ of electrons at the surface vs. temperature $T$. Inset: diffusion coefficient $D_2$ of electrons at the surface vs. temperature $T$. (d) Dephasing lengths $L_{\varphi_1}$ and $L_{\varphi_2}$ vs. temperature $T$. Both $L_{\varphi_1}$ and $L_{\varphi_2}$ are found to exhibit a power-law temperature dependence as $L_{\varphi_1} \sim T^{-0.55}$ and $L_{\varphi_2} \sim T^{-0.49}$. (e) $\tau_{\varphi_1}/\tau_{SB_1}$ vs. temperature $T$. (f) $\tau_{\varphi_2}/\tau_{SB_2}$ vs. temperature $T$. Here, we note that $\tau_{\varphi_2}/\tau_{SB_2} = 2.9$ at $T = 14$ K.

# Supplementary Materials for
# Surface-bulk coupling in a Bi$_2$Te$_3$ nanoplate grown by van der Waals epitaxy

Xiaobo Li[1,2], Mengmeng Meng[1], Shaoyun Huang[1], Congwei Tan[3], Congcong Zhang[3], Hailin Peng[3], and H. Q. Xu[1,4,*]

[1] *Beijing Key Laboratory of Quantum Devices, Key Laboratory for the Physics and Chemistry of Nanodevices, and School of Electronics, Peking University, Beijing 100871, China*

[2] *Academy for Advanced Interdisciplinary Studies, Peking University, Beijing 100871, China*

[3] *Center for Nanochemistry, Beijing National Laboratory for Molecular Sciences (BNLMS), College of Chemistry and Molecular Engineering, Peking University, Beijing 100871, China*

[4] *Beijing Academy of Quantum Information Sciences, Beijing 100193, China*

[*] Corresponding authors: H. Q. Xu (hqxu@pku.edu.cn)

(Date: January 19, 2022)

## Contents

**Supplementary Note I. Characteristic transport parameters that would be obtained by fits of the measured magnetoconductivity data to the Hikami-Larkin-Nagaoka (HLN) formula**

**Supplementary Note II. Surface-bulk inter-channel scattering rates, $\tau_{SB_1}^{-1}$ and $\tau_{SB_2}^{-1}$, and the diffusion coefficients in the bulk and at the top surface, $D_1$ and $D_2$**

**Supplementary note I. Characteristic transport parameters that would be obtained by fits of the measured magnetoconductivity data to the Hikami-Larkin-Nagaoka (HLN) formula**

In this supplementary note, we show the fits of the measured magnetoconductivity data at different top-gate voltages $V_g$ and temperatures $T$ based on the Hikami-Larkin-Nagaoka (HLN) theory. The HLN theory does not take inter-channel scattering into account. For a two-dimensional (2D) disordered conducting system, the theory predicts that the quantum correction to the classical conductivity can be written as[1]

$$\Delta\sigma_{xx}(B) = -\alpha \frac{e^2}{\pi h}\left[\Psi\left(\frac{\hbar}{4eBL_\varphi^{*2}}+\frac{1}{2}\right) - ln\left(\frac{\hbar}{4eBL_\varphi^{*2}}\right)\right], \qquad (1)$$

where prefactor $\alpha$ refers to the number of 2D conduction channels in the system, with $\alpha \sim 1/2$ representing one 2D conduction channel and $\alpha \sim 1$ representing two separated 2D conduction channels, and $L_\varphi^*$ is the dephasing length of electrons in the system. To clearly present the fits of our experimental data to the HLN theory, we show in Fig. S1(a) again the magnetoconductivity data (opened circles) measured for the device at different gate voltages $V_g$ at $T = 2$ K as in Fig. 2(a) of the main article. Again, the black open circles are obtained at $V_g = 11$ V and the ones at other values of $V_g$ are successively vertically offset for clarify. Opened circles in Fig. S1(c) show the same magnetoconductivity data measured for the device at different temperatures $T$ at $V_g = -10$ V as in Fig. 3(a) of the main article. Here, the dark green circles are obtained at $T = 2$ K and the ones at other values of $T$ are again successively vertically offset for clarify. The dashed lines in Figs. S1(a) and S1(c) show the fits of the measured magnetoconductivity data based on the HLN theory. Figure S1(b) displays the extracted $L_\varphi^*$ and $\alpha$ from the fits shown in Fig. S1(a), while Fig. S1(d) displays the extracted $L_\varphi^*$ and $\alpha$ from the fits shown in Fig. S1(c). It is seen that both $L_\varphi^*$ and $\alpha$ are top-gate voltage or temperature dependent. The extracted value of $\alpha$ increases with decreasing top-gate voltage or increasing temperature, reflecting that the electron transport at the 2D top surface channel and in the 2D bulk channel is gradually coherently decoupled in the $Bi_2Te_3$ nanoplate. The dephasing length $L_\varphi^*$ is shown to decrease as the top-gate voltage decreases, which is a result of an enhancement in electron-electron interaction[2] at a lower electron density. These results are qualitatively in agreement with the results obtained based on the Garate-Glazman theory as presented in the main article. However, the extracted $L_\varphi^*$ exhibits a power-law temperature dependence, $L_\varphi^* \sim T^{-0.38}$. This power-law temperature dependence is incompatible with the well-established dephasing process caused by electron-electron scattering with small-energy transfers in a 2D system at low temperatures.

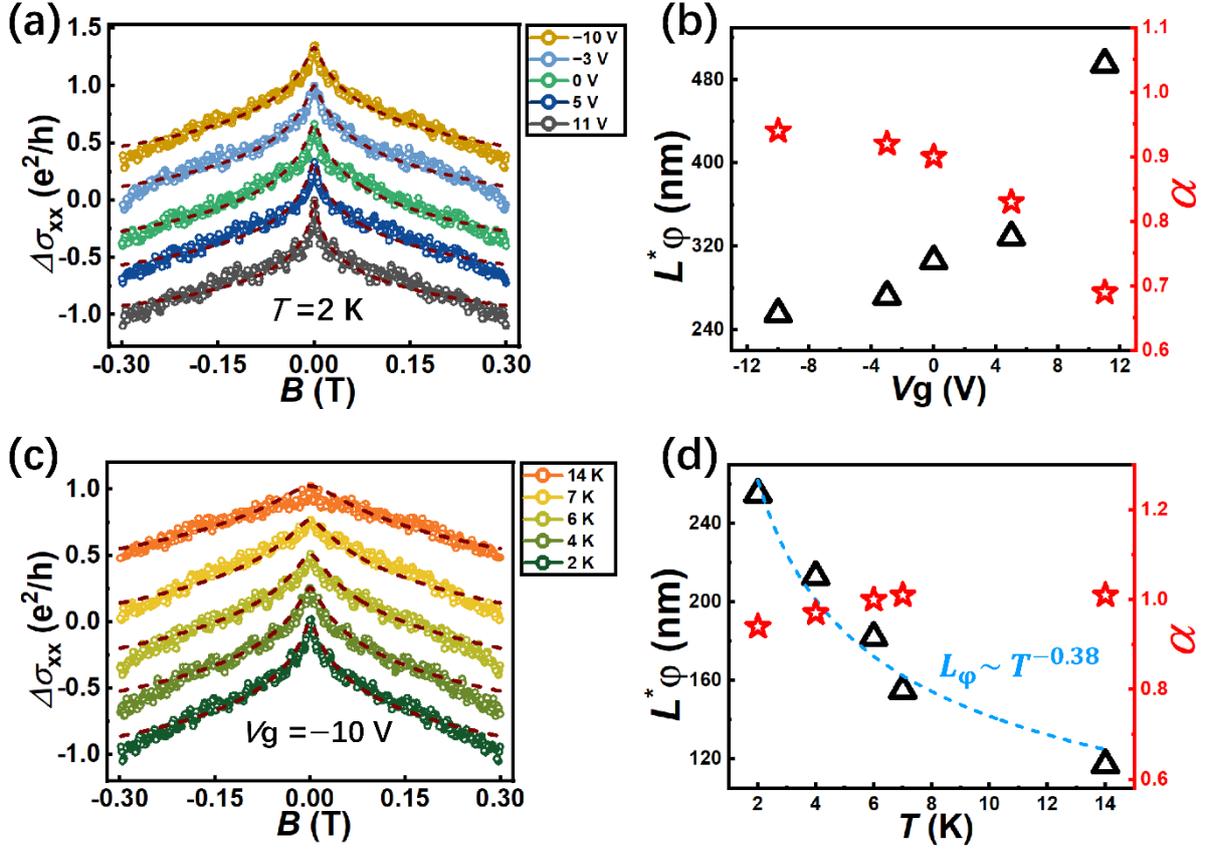

FIG. S1. (a) Magnetoconductivity $\Delta\sigma_{xx}$ of the device at different top-gate voltages at $T = 2$ K. The opened circles show the measured data. The black ones are obtained at $V_g = 11$ V and the ones at other top-gate voltages are successively vertically offset for clarify. The dashed lines show the fits of the measured data based on the HLN theory. (b) Extracted dephasing length $L_\varphi^*$ and prefactor $\alpha$ vs. top-gate voltage $V_g$. (c) Magnetoconductivity $\Delta\sigma_{xx}$ of the device at different temperatures $T$ at $V_g = -10$ V. Again, the opened circles show the measured data. The dark green ones are obtained at $T = 2$ K and the ones at other temperatures are successively vertically offset for clarify. The dashed lines show the fits of the measured data based on the HLN theory. (d) Extracted dephasing length $L_\varphi^*$ and prefactor $\alpha$ vs. temperature $T$.

**Supplementary note II. Surface-bulk inter-channel scattering rates, $\tau_{SB_1}^{-1}$ and $\tau_{SB_2}^{-1}$, and the diffusion coefficients in the bulk and at the top surface, $D_1$ and $D_2$**

In this supplementary note, we discuss how the surface-bulk inter-channel scattering rates, $\tau_{SB_1}^{-1}$ and $\tau_{SB_2}^{-1}$, and the diffusion coefficients, $D_1$ and $D_2$, are changed with changing top-gate voltage $V_g$. In the simple Fermi golden role approximation, the bulk-to-surface electron inter-channel scattering rate is proportional to the density of states (DOS) at the top surface, $\tau_{SB_1}^{-1}(E_F) \propto \mathrm{DOS}_2(E_F)$, where the density of states at the top surface $\mathrm{DOS}_2(E_F)$ is given by

$$\mathrm{DOS}_2(E_F) = \frac{E_F}{\pi \hbar^2 v_F^2} \ , \tag{2}$$

with $\hbar$ being the reduced Planck Constant, $E_F$ the top-surface electron Fermi energy measured with respect to the Dirac point, and $v_F$ the electron Fermi velocity at the top surface. It is seen that $\mathrm{DOS}_2(E_F)$ is a linear function $E_F$ and is thus decreased with decreasing $V_g$. As a result, the bulk-to-surface electron inter-channel scattering rate $\tau_{SB_1}^{-1}$ is decreased with decreasing $V_g$, leading to an increase in the bulk electron mobility with decreasing $V_g$ at low temperatures.

The surface-to-bulk electron scattering rate is proportional to the DOS in the bulk, $\tau_{SB_2}^{-1}(E_F) \propto \mathrm{DOS}_1(E_F)$, where the DOS in the bulk $\mathrm{DOS}_1(E_F)$ is given by

$$\mathrm{DOS}_1(E_F) = \frac{m^*}{\pi \hbar^2} \ , \tag{3}$$

with $E_F$ being the bulk electron Fermi energy measured with respect to the bulk conduction band bottom and $m^*$ the electron effective mass in the bulk. Here, in deriving $\mathrm{DOS}_1(E_F) = \frac{m^*}{\pi \hbar^2}$, we assume there is only one 2D subband in the bulk. Evidently, since $\mathrm{DOS}_1(E_F)$ is independent of the Fermi energy, the surface-to-bulk electron inter-channel scattering rate $\tau_{SB_2}^{-1}$ is $V_g$-independent, which could imply that the surface-to-bulk electron scattering would not give a visible change in the surface electron mobility with a change in $V_g$.

Now we discuss the dependences of the diffusion coefficients $D_1$ and $D_2$ on the top-gate voltage $V_g$. Based on the Einstein relation for the conductivity $\sigma$ and the diffusion coefficient $D$ in a conductor, $\sigma = e^2 \mathrm{DOS}(E_F) D$, the diffusion coefficient $D$ of electrons in the conductor can be expressed as $D = \frac{n\mu}{e\mathrm{DOS}(E_F)}$, where $n$ and $\mu$ are the electron density and mobility in the conductor, and $e$ the elementary charge. More specifically, for the 2D bulk channel in our $Bi_2Te_3$ nanoplate (with the occupation of the lowest 2D subband assumed), the electron density $n_1(E_\text{F})$ and the density of states $\text{DOS}_1(E_\text{F})$ are related by $n_1(E_\text{F})/\text{DOS}_1(E_\text{F}) = E_\text{F}$. Thus, the diffusion coefficient $D_1$ of electrons in the bulk is expressed as

$$D_1 = \frac{\mu_1 E_\text{F}}{e} , \tag{4}$$

where $\mu_1$ denotes the electron mobility in the bulk. For the 2D top surface channel, the electron density $n_2(E_\text{F})$ and the density of states $\text{DOS}_2(E_\text{F})$ are related by $n_2(E_\text{F})/\text{DOS}_2(E_\text{F}) = E_\text{F}/2$. The diffusion coefficient $D_2$ of electrons at the top surface can then be expressed as

$$D_2 = \frac{\mu_2 E_\text{F}}{2e} , \tag{5}$$

where $\mu_2$ denotes the electron mobility at the top surface.

The above analyses show that the diffusion coefficients $D_1$ and $D_2$ in the bulk and at the top surface could be expressed as a function of a product of the mobility and the Fermi energy of electrons in their respective subsystems. In the experiment, each mobility value of $\mu$ shown in the inset of Fig. 1(c) of the main article is in fact an averaged value over the electrons at the top surface and in the bulk and is an outcome of the presence of all scattering events (including both intra-channel and inter-channel scatterings). If we take the experimentally measured $\mu$ for both $\mu_1$ and $\mu_2$, which is increased with decreasing $V_\text{g}$ [see the inset of Fig. 1(c) of the main article], and consider the fact that the Fermi energies $E_\text{F}$ at the top surface and in the bulk are decreased with decreasing $V_\text{g}$, we expect that both $D_1$ and $D_2$ would exhibit a weak $V_\text{g}$-dependence, which is consistent with our experimental observations shown in the inset of Figs. 2(b) and 2(c) of the main article.